\documentclass[10pt,letterpaper]{article}
\usepackage[top=0.85in,left=2.75in,footskip=0.75in]{geometry}

\usepackage{amsmath,amssymb}

\usepackage{changepage}

\usepackage{textcomp,marvosym}
\usepackage[utf8]{inputenc}
\usepackage[T1]{fontenc}
\usepackage{float}
\usepackage{algorithm}
\usepackage{algpseudocode}
\usepackage{float}
\usepackage{tabularx}
\usepackage{booktabs}
\usepackage{cite}

\usepackage{nameref,hyperref}

\usepackage[right]{lineno}

\usepackage[nopatch=eqnum]{microtype}
\DisableLigatures[f]{encoding = *, family = * }

\usepackage[table]{xcolor}

\usepackage{array}

\newcolumntype{+}{!{\vrule width 2pt}}

\newlength\savedwidth

\raggedright
\usepackage[aboveskip=1pt,labelfont=bf,labelsep=period,justification=raggedright,singlelinecheck=off]{caption}

\makeatletter
\renewcommand{\@biblabel}[1]{\quad#1.}
\makeatother

\usepackage{lastpage,fancyhdr,graphicx}
\usepackage{epstopdf}
\fancyheadoffset[L]{2.25in}
\fancyfootoffset[L]{2.25in}
\begin{document}
\vspace*{0.2in}

\begin{flushleft}
{\Large
\textbf\newline{Predicting ICU Readmission in Acute Pancreatitis Patients Using a Machine Learning-Based Model with Enhanced Clinical Interpretability} 
}
\newline
\\
Shuheng Chen\textsuperscript{1},
Yong Si\textsuperscript{1},
Junyi Fan\textsuperscript{1},
Li Sun\textsuperscript{1},
Elham Pishgar\textsuperscript{2},
Kamiar Alaei\textsuperscript{3},
Greg Placencia\textsuperscript{4},
Maryam Pishgar\textsuperscript{1*}
\textsuperscript{\textpilcrow}
\\
\bigskip
\textbf{1} Department of Industrial and Systems Engineering, University of Southern California, 3715 McClintock Ave GER 240, Los Angeles, 90087, California, United States
\\
\textbf{2} Colorectal Research Center, Iran University of Medical Sciences, Tehran Hemat Highway next to Milad Tower, Tehran, 14535, Iran
\\
\textbf{3} Department of Health Science, California State University, Long Beach (CSULB), 1250 Bellflower Blvd, Long Beach, 90840, California, United States
\\
\textbf{4} Department of Industrial and Manufacturing Engineering, California State Polytechnic University, Pomona, 3801 W Temple Ave, Pomona, 91768, California, United States
\\
\bigskip

%
%




\textpilcrow Membership list can be found in the Acknowledgments section.

* pishgar@usc.edu

\end{flushleft}
\section*{Abstract}
Acute pancreatitis (AP) is a common and potentially life-threatening gastrointestinal disease that places a substantial burden on healthcare systems worldwide. ICU readmissions among patients with AP remain frequent, especially in severe or recurrent cases, with rates exceeding 40\%. Timely identification of patients at high risk for readmission is critical for guiding clinical decision-making and improving outcomes. In this study, we used the MIMIC-III database to identify ICU admissions for AP based on standardized diagnostic codes.

We implemented a structured preprocessing pipeline that included missing data imputation, correlation analysis, and hybrid feature selection. Specifically, we applied Recursive Feature Elimination with Cross-Validation (RFECV) and LASSO regression, supported by clinical expert review, to reduce an initial set of over 50 variables to 20 key predictors encompassing demographics, comorbidities, laboratory tests, and interventions. To address class imbalance, the Synthetic Minority Over-sampling Technique (SMOTE) was incorporated within a stratified five-fold cross-validation framework to maintain balanced training and unbiased evaluation.

Six machine learning models—Logistic Regression, k-Nearest Neighbors, Naive Bayes, Random Forest, LightGBM, and XGBoost—were developed and optimized through grid search. Model performance was assessed using standard metrics including AUROC, accuracy, F1 score, sensitivity, specificity, Positive Predictive Value (PPV), and Negative Predictive Value (NPV). XGBoost achieved the best performance, with an AUROC of 0.862 (95\% CI: 0.800–0.920) and accuracy of 0.889 (95\% CI: 0.858–0.923) on the test set.

An ablation study demonstrated the importance of each selected feature, as removing any one led to a reduction in model performance. Furthermore, SHAP (SHapley Additive exPlanations) analysis was conducted to enhance interpretability. Platelet count, age, and peripheral oxygen saturation (SpO$_2$) were identified as major contributors to readmission prediction. Overall, this study shows that ensemble learning, informed feature selection, and class imbalance handling can improve prediction of ICU readmission risk in patients with AP. These findings may support the development of more targeted post-discharge interventions to reduce preventable readmissions.


\section*{Introduction}
Acute pancreatitis (AP) is a prevalent and potentially life-threatening gastrointestinal disorder, accounting for over 275,000 hospital admissions annually in the United States, with associated healthcare costs exceeding \$2.6 billion per year\cite{fagenholz2007direct}. Globally, its incidence ranges from 5 to 80 cases per 100,000 population, with rising trends observed over the past decade\cite{banks2013classification,peery2012burden,roberts2008incidence,xiao2016global}. AP disproportionately affects adults aged 30–60 years, though cases span all age groups, and alcohol-related AP is more common in males, while gallstone-induced AP predominates in
females\cite{yadav2013epidemiology,spanier2008epidemiology}. Severe AP carries a mortality rate of 10\%–30\%, underscoring its significant burden on healthcare systems\cite{fagenholz2007direct,banks2013classification,van2017acute}. 

AP readmissions in Intensive Care Unit (ICU) constitute a major healthcare challenge, imposing substantial clinical and economic burdens on healthcare systems globally. Epidemiological studies demonstrate that 15-25\% of AP patients require rehospitalization within 30 days, with rates exceeding 40\% among those with severe or recurrent disease\cite{krishna2017changing,munigala2017predictors,peery2025burden}. The persistent morbidity and healthcare utilization associated with AP readmissions highlight critical gaps in post-discharge management and transitional care, despite established guideline recommendations\cite{yadav2014population,garg2018incidence}. Patients with severe AP face particularly elevated risks, with readmissions predominantly attributable to three key factors: (1) persistent multi-organ dysfunction (especially renal and respiratory failure), (2) infectious complications including infected pancreatic necrosis, and (3) progression of local pancreatic complications such as walled-off necrosis or pseudocysts\cite{whitlock2010early,garg2018incidence}. The ability to predict readmission risk in AP is thus a crucial factor in optimizing long-term outcomes, as timely interventions may mitigate preventable complications and reduce healthcare costs.

Whitlock et al. (2011) developed a risk prediction scoring system aimed at assessing the likelihood of ICU readmission in patients diagnosed with AP.\cite{whitlock2011scoring} This model incorporated critical discharge characteristics, including gastrointestinal symptoms, intolerance to solid food, pancreatic necrosis, antibiotic use, and persistent pain. The study demonstrated that the scoring system effectively stratified patients into three distinct risk categories: low, moderate, and high. The corresponding readmission rates for these groups were 4\%, 15\%, and 87\%, respectively. Furthermore, the model exhibited strong predictive performance, as evidenced by an area under the receiver operating characteristic curve (AUROC) of 0.83 in a validation cohort.

Buxbaum et al. (2018) assessed the utility of the Pancreatic Activity Scoring System (PASS) in predicting ICU readmission risk for patients with AP.\cite{buxbaum2018pancreatitis} The PASS includes factors such as organ failure, intolerance to solid food, systemic inflammatory response syndrome (SIRS), pain intensity, and opioid use. The study demonstrated that a PASS score above 60 significantly increased the likelihood of readmission, with a five-fold higher risk compared to patients with lower scores. The sensitivity and specificity of PASS for predicting readmission were found to be 68\% and 71\%, respectively, with an AUROC of 0.750.

Ding et al. (2021) proposed a nomogram designed to predict the risk of 30-day ICU readmission in patients with AP.\cite{ding2021nomogram} The model was constructed using several clinical variables, including etiology (biliary or alcoholic), infected pancreatic necrosis, total bilirubin levels, blood glucose concentration, and serum albumin levels. The performance of the nomogram was evaluated in terms of its sensitivity and specificity, which were found to be 66.7\% and 75.4\%, respectively. Additionally, the nomogram demonstrated a satisfactory predictive ability, with an AUROC of 0.780, suggesting its potential clinical utility in risk assessment.

While traditional models like nomograms and the PASS scoring system have provided useful insights for predicting readmission in acute pancreatitis, they have limitations. These models often rely on a limited set of variables, which may not fully capture the complexity of individual patients. Additionally, they can struggle with high-dimensional data or missing values, potentially impacting their generalizability and predictive accuracy. Accordingly, there is a pressing need for a novel predictive tool capable of addressing the limitations of existing models, while offering accurate and robust predictions of readmission risk in ICU patients with AP.

Recent years have witnessed significant progress in machine learning applications for clinical outcome prediction. Especially, XGBoost (eXtreme Gradient Boosting) has emerged as a particularly effective algorithm due to its three distinctive advantages: (1) built-in regularization that prevents overfitting to medical datasets with high feature dimensionality, (2) native support for handling missing values common in electronic health records, and (3) superior feature importance quantification that aligns with clinical interpretability needs.\cite{chen2016xgboost} For example, Chen et al. (2025) developed an XGBoost-based model to predict ICU mortality in sepsis-associated acute kidney injury patients using the MIMIC-IV and eICU databases. The model demonstrated strong discriminative performance, achieving an AUROC of 0.878 (95\% Confidence Interval (CI) : 0.859–0.897) in internal validation, with further confirmation in an external validation cohort.\cite{chen2025xgboost} Ashrafi et al. (2024) developed an XGBoost-based predictive model for mortality risk assessment in ICU-admitted heart failure patients, achieving an AUROC of 0.923 (95\% CI: 0.875–0.961). Their findings underscore the strong discriminative capability of XGBoost in modeling critical care outcomes, reinforcing its utility as a high-performance machine learning approach for clinical decision support in intensive care settings.\cite{ashrafi2024optimizing}

This study introduces several methodological innovations that enhance the predictive performance and clinical applicability of machine learning models for ICU readmission risk in AP patients. 
\begin{itemize}
    \item An integrated feature selection strategy was employed to refine an initial pool of 50 candidate variables down to 20 key predictors. This process combined Recursive Feature Elimination with Cross-Validation (RFECV) and Least Absolute Shrinkage and Selection Operator (LASSO) regression, with final feature retention guided by expert clinical judgment.
    \item To address class imbalance, the study employs the Synthetic Minority Over-sampling Technique (SMOTE) within a rigorous 5-fold cross-validation framework, thereby improving the robustness and generalizability of the predictive model.
    \item A thorough evaluation of the proposed model, encompassing an in-depth analysis of the confusion matrix, AUROC curves, and a structured ablation study, revealed notable performance gains. The model achieved an AUROC of 0.862 (95\% CI: 0.800–0.920) and an accuracy of 0.889 (95\% CI: 0.858–0.923), surpassing the performance reported in prior studies.
    \item To enhance model interpretability and support its potential clinical utility, we conducted a SHapley Additive exPlanations (SHAP) analysis to quantify the contribution of individual features. Key variables such as platelet count, patient age, and peripheral oxygen saturation (SpO$_2$) were identified as prominent contributors to model predictions.
\end{itemize}

\section*{Methods}
\subsection*{Data Source and study design}

This study utilized the Medical Information Mart for Intensive Care III (MIMIC-III), a widely used, publicly available critical care database developed through a joint initiative between the Beth Israel Deaconess Medical Center (BIDMC) and the Massachusetts Institute of Technology (MIT). MIMIC-III contains de-identified, high-resolution clinical data from patients admitted to ICU, and has been extensively validated in prior research. The comprehensive nature of the dataset provides a reliable foundation for developing and evaluating predictive models in critical care settings.

The database includes a wide spectrum of information such as patient demographics, vital signs, laboratory results, medication administration, fluid balance, and outcomes. It supports standardized disease classification using both ICD-9 and ICD-10 codes. A notable feature of MIMIC-III is the availability of high-resolution, hourly physiological data from bedside monitors, verified by ICU professionals. The breadth and granularity of this dataset make it particularly well-suited for developing and validating machine learning models in critical care research.

To systematically represent the analytical process of this study, we present a rule-based pseudocode workflow in Algorithm~\ref{alg:readmission}. This pipeline captures the entire machine learning process, including data selection, preprocessing, feature engineering, feature selection, model training, statistical validation, and interpretability analysis. Each step is executed in a modular and reproducible manner, enabling robustness and clinical applicability. This structured representation facilitates reproducibility, algorithmic transparency, and implementation consistency.

\begin{algorithm}[H]
\caption{\textbf{ML Pipeline for 7-Day ICU Readmission Risk in Acute Pancreatitis}}
\label{alg:readmission}
\begin{algorithmic}[1]
\Require MIMIC-III ICU patient data diagnosed with acute pancreatitis
\Ensure Readmission risk prediction within 7 days

\State \textbf{Step 1: Patient Selection and Filtering}
\State Identify AP patients using ICD-9/10 codes
\State Exclude: age < 18, ICU stay < 24h, renal history
\State Retain first ICU admission per patient

\State \textbf{Step 2: Data Preprocessing}
\ForAll{feature in dataset}
    \If{feature is numeric}
        \State compute \texttt{min}, \texttt{max}, and \texttt{mean} values
    \ElsIf{feature is categorical}
        \State apply label encoding
    \EndIf
\EndFor
\State Impute numeric missing values with median
\State Impute categorical missing values with mode
\State Normalize all continuous variables via min-max scaling

\State \textbf{Step 3: Feature Selection}
\State Apply RFECV based on Gini importance (Random Forest)
\State Apply LASSO regression to shrink less relevant variables
\State Retain intersection of top-ranked features

\State \textbf{Step 4: Data Balancing}
\State Use SMOTE to oversample minority class (readmitted cases)

\State \textbf{Step 5: Model Development and Tuning}
\State Split data into training (70\%) and test (30\%) sets
\ForAll{model $\in$ \{XGBoost, Random Forest, LightGBM, AdaBoost, LR, NN\}}
    \State Conduct GridSearchCV for hyperparameter tuning
    \State Evaluate model on AUROC, accuracy, precision, recall, F1
\EndFor

\State \textbf{Step 6: Statistical and Interpretability Analysis}
\State Perform t-test and chi-square test on training vs. test cohorts
\State Apply SHAP to interpret top predictors
\State Conduct ablation study to evaluate feature contribution

\State \textbf{Step 7: Final Validation}
\State Report AUROC with 95\% CI using 2000 bootstrap replicates
\State Visualize calibration curve and decision threshold
\end{algorithmic}
\end{algorithm}

\subsection*{Patient Selection}

Patients diagnosed with AP were identified from the MIMIC-III\cite{johnson2016mimic} database using standardized diagnostic codes. Specifically, inclusion criteria were based on the International Classification of Diseases, Ninth Revision (ICD-9) code 577.0, which corresponds to AP. An initial cohort of 1,619 patients was obtained. Data extraction was performed using Structured Query Language (SQL) within a PostgreSQL environment to ensure reproducibility and data integrity.

Subsequent exclusion criteria were systematically applied to enhance cohort consistency and clinical relevance. First, 219 patients were excluded due to age younger than 18 years. Second, 286 patients were excluded because their ICU stay duration was less than 24 hours. Third, 31 patients with a documented history of renal disease were removed. To prevent duplication bias, for patients with multiple ICU admissions, only data from the first ICU stay were retained, resulting in the removal of an additional 0 patients.

After all exclusions, a final cohort of 1,083 patients was identified for analysis. For each included patient, comprehensive clinical data were extracted, including demographics, comorbidity profiles, vital signs, laboratory measurements, and interventions such as mechanical ventilation, renal replacement therapy, and administration of vasoactive agents. All vital signs and laboratory values were obtained from the first 24 hours following ICU admission to capture the initial physiological status.

The primary clinical endpoint was defined as ICU readmission within seven days of discharge from the index ICU stay. Readmission was characterized according to widely accepted ICU quality metrics: an unplanned return to the ICU within 48 to 72 hours after discharge, or any subsequent ICU admission during the same hospitalization period \cite{ruppert2023predictive, ouanes2012model}. Planned step-down unit transfers or scheduled ward movements were excluded. This methodology ensures that the final study cohort possesses complete, high-quality information crucial for robust modeling of early ICU readmission risk among AP patients.

\begin{figure}[H]
\centering
\includegraphics[width=1\linewidth]{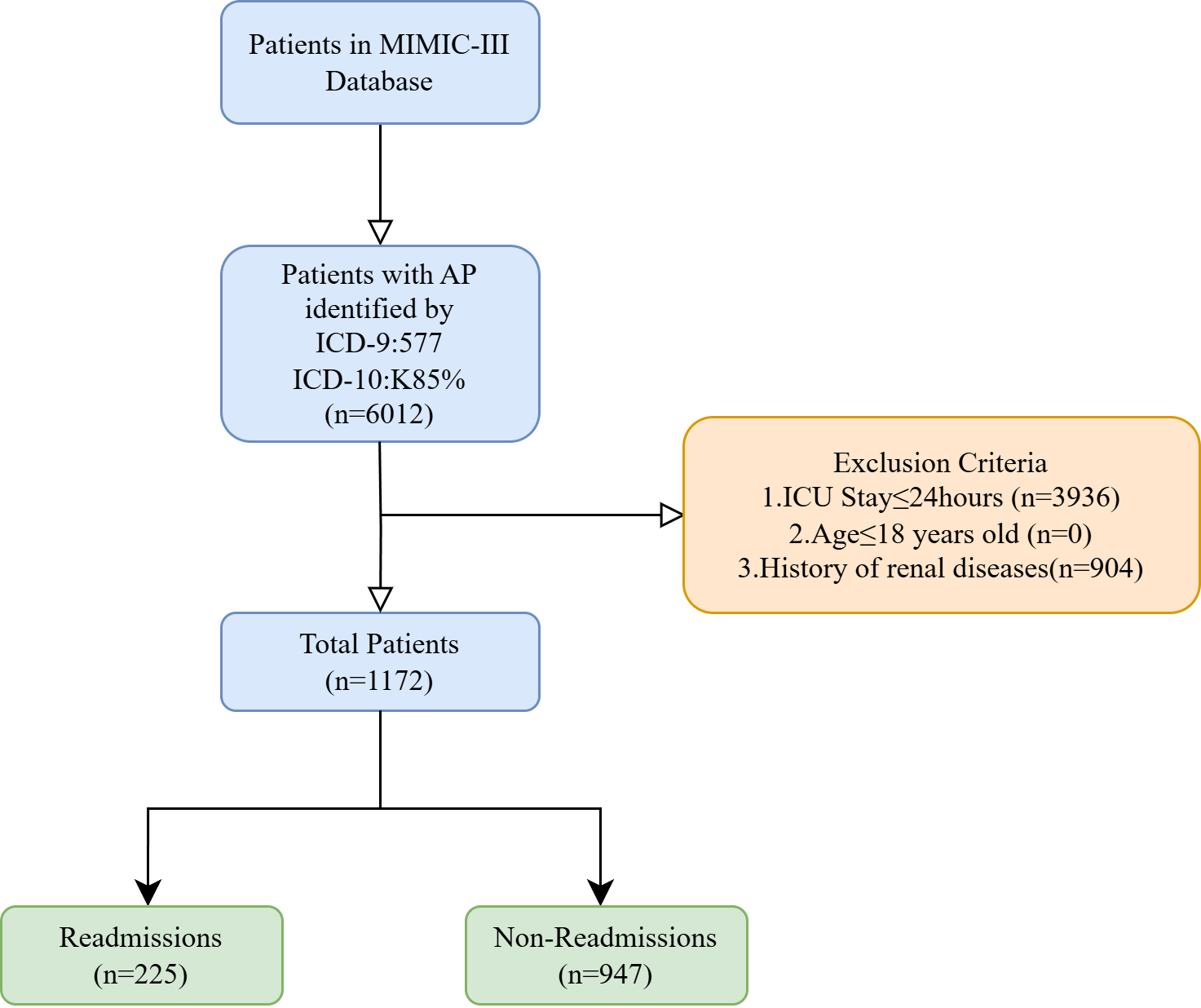} 
\caption{\textbf{Flowchart Illustrating the Patient Selection Process from MIMIC-III}}
\label{fig:patient_selection}
\end{figure}

The patitent selection process is illustrated in Figure \ref{fig:patient_selection}

\subsection*{Data Preprocessing}

In the data preprocessing phase, we employed a systematic and multi-step strategy to ensure the quality, consistency, and analytical robustness of the dataset for subsequent modeling. Each operation was designed not only to address common issues in real-world clinical data, such as missingness and heterogeneity, but also to align with the assumptions and requirements of downstream machine learning algorithms.

First, missing values were handled separately for numerical and categorical features to minimize bias. For numerical variables, missing entries were imputed using the median, a strategy known for its robustness against outliers and skewed distributions, thus preserving the central tendency of the original data \cite{Si2025.03.14.25324005, fan2025lightgbm}. For categorical features, missing values were imputed using the mode, ensuring consistency by replacing gaps with the most representative category.

Following imputation, new derived variables were created to capture dynamic trends during the first 24 hours of ICU stay. Specifically, for selected vital signs and laboratory indicators (e.g., glucose, blood pressure), we computed the minimum, maximum, and mean values, thereby enriching the feature space with temporal information. These derived statistics were formalized as:

\begin{equation}
x_{i,\text{mean}} = \frac{1}{T} \sum_{t=1}^{T} x_{i}^{(t)}, \quad 
x_{i,\text{max}} = \max_{t \in [1,T]} x_{i}^{(t)}, \quad 
x_{i,\text{min}} = \min_{t \in [1,T]} x_{i}^{(t)}
\label{eq:temporal_features}
\end{equation}

This step allowed the model to capture both static snapshots and short-term variability in patient physiology, which are clinically meaningful predictors of acute outcomes.

To further prepare the data for machine learning modeling, continuous variables were scaled to a uniform 0--1 range using min-max normalization. This rescaling was crucial to prevent features with larger magnitudes from disproportionately influencing the model learning process, particularly for algorithms sensitive to feature scales, such as gradient boosting frameworks.

Additionally, patients who failed to meet the predefined clinical inclusion criteria (e.g., ICU stay duration, age thresholds, renal disease history) were excluded at this stage, ensuring the homogeneity and clinical relevance of the final analytical cohort.

To address the class imbalance inherent in the ICU readmission dataset, we applied SMOTE within a 5-fold cross-validation framework. SMOTE was applied exclusively to the training folds, where it synthetically generated new minority class instances by interpolating between existing samples, thereby enhancing the representation of positive (readmission) cases. Importantly, the validation folds remained untouched to preserve the original class distribution, ensuring that performance metrics reflected the model's ability to generalize to real-world, imbalanced data. This strategy not only improved the sensitivity of the predictive models toward the minority class but also minimized the risk of data leakage, as all preprocessing steps—including imputation, scaling, and feature encoding—were fitted solely on the training partitions and subsequently applied to the validation partitions. Following this procedure, final model evaluation was conducted on a holdout test set that remained completely isolated throughout training and tuning phases. This rigorous methodology enhanced the robustness, generalizability, and credibility of our predictive modeling results.

Overall, the preprocessing pipeline was designed to maximize data integrity while preserving clinically important signals, enabling robust, unbiased, and interpretable modeling in the prediction of ICU readmission risk among critically ill patients with acute pancreatitis.

\subsection*{Feature Selection}

\begin{figure}[htbp]
\centering
\includegraphics[width=1\linewidth]{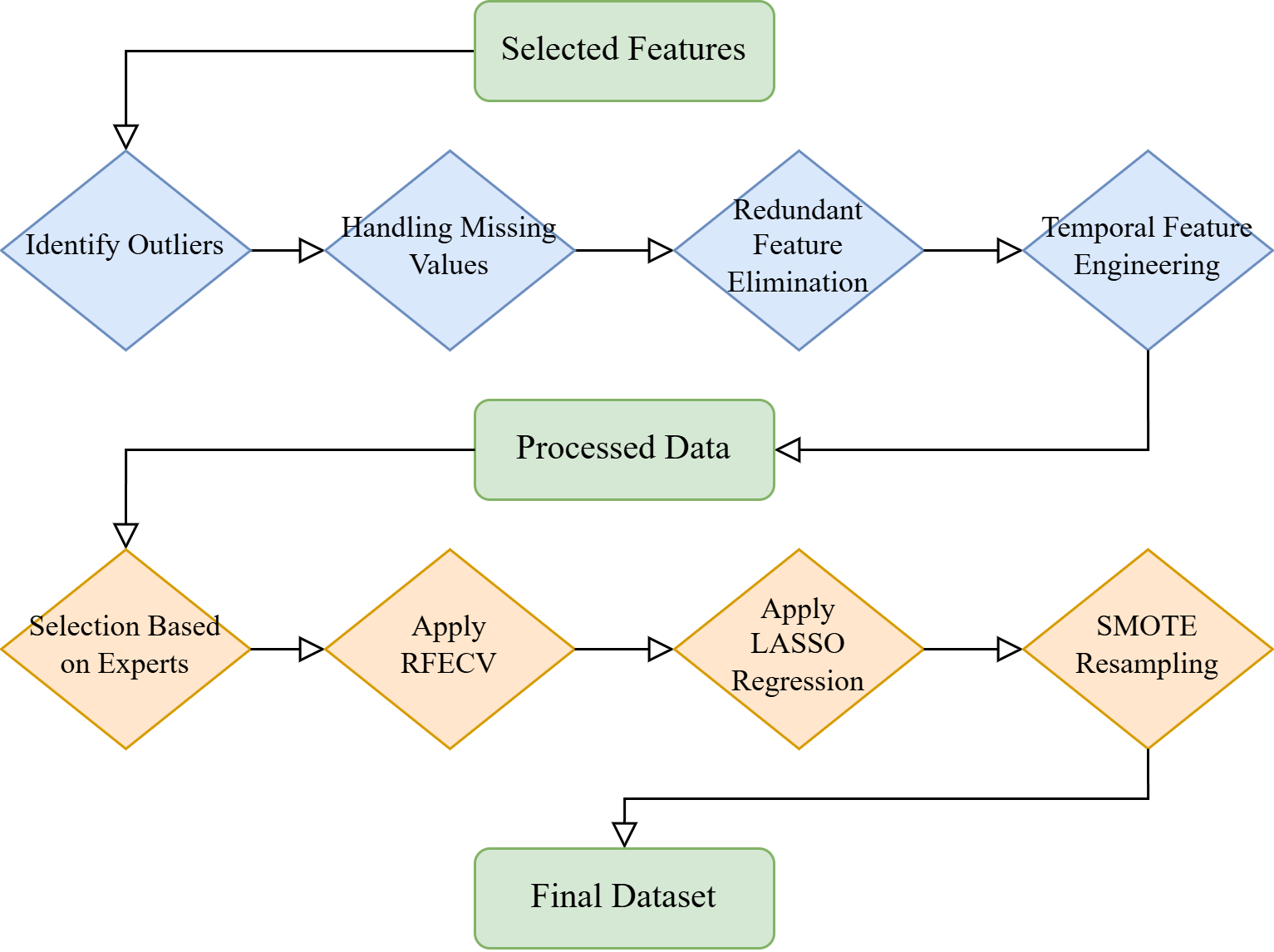}
\caption{\textbf{Flowchart Illustrating the Full Data Preprocessing and Feature Selection Pipeline.}}
\label{fig:preprocessing_flow}
\end{figure}

A hybrid feature selection strategy was adopted to identify the most informative variables for predicting ICU readmission in patients with AP. This approach combined both RFECV \cite{guyon2002gene} and LASSO \cite{tibshirani1996regression} techniques, which have been widely validated in clinical predictive modeling \cite{chen2023lightgbm}. The full data preprocessing and feature selection workflow is illustrated in Figure~\ref{fig:preprocessing_flow}.

 In our study, the 50 initial candidate features were categorized into five primary domains based on their clinical significance and contribution to patient profiling. (1) For demographic characteristics, features such as age, gender, insurance type, and ethnicity were included to account for potential population heterogeneity and socioeconomic factors that may influence ICU outcomes. (2) Regarding vital signs, we incorporated early measurements such as heart rate, systolic and diastolic blood pressure, mean arterial pressure, respiratory rate, SpO$_2$, and body temperature, all recorded within the first 24 hours of ICU admission, reflecting acute physiological status. (3) In terms of laboratory results, comprehensive biochemical and hematological markers were collected, including serum glucose, electrolytes (sodium, potassium, chloride), renal function indicators (creatinine, blood urea nitrogen), coagulation parameters (prothrombin time, INR), and complete blood counts, providing objective insights into systemic organ function. (4) Comorbidity features, such as the presence of cardiovascular disease, diabetes mellitus, chronic respiratory conditions, dementia, malignancy, and sepsis, were included to capture baseline disease burdens known to affect critical care trajectories. (5) Finally, clinical interventions and outcome-related features were considered, including ICU length of stay, hospital length of stay, use of mechanical ventilation, and renal replacement therapy, representing the intensity and complexity of care each patient received during their hospitalization. This structured categorization provided a comprehensive and clinically interpretable feature space for subsequent predictive modeling.

The selection of these candidate features was meticulously guided by consultations with a clinical expert and a comprehensive review of prior literature\cite{harutyunyan2019multitask,johnson2016machine}. This rigorous approach ensured that each variable retained for initial consideration was highly relevant to the clinical questions posed, thereby significantly enhancing the robustness, clinical interpretability, and practical applicability of our predictive models. This thoughtful compilation of features enabled a holistic representation of patient health status, supporting the identification of clinically meaningful patterns and predictors for ICU readmission among critically ill patients with acute pancreatitis.
Following the initial feature compilation, a hybrid feature selection strategy was adopted to systematically refine the feature set and enhance model parsimony without sacrificing predictive performance. Specifically, RFECV and LASSO regression were employed to identify and retain the most informative variables. These complementary techniques leveraged both model-driven and statistical criteria to optimize feature relevance, generalizability, and model interpretability, as detailed in subsequent sections.

First, RFECV—a wrapper-based method—was applied to iteratively remove the least important features while using stratified cross-validation to determine the optimal subset that maximized predictive performance. In each iteration, feature importance was assessed using the Gini Importance metric from a Random Forest classifier, defined as:

\begin{equation}
I(x_i) = \sum_{t \in T} \frac{p(t) \cdot \Delta i(t)}{f(t)}
\label{eq:gini}
\end{equation}

where $I(x_i)$ denotes the importance score of feature $x_i$, $T$ is the set of all decision trees in the random forest, $p(t)$ is the proportion of samples reaching node $t$, $\Delta i(t)$ is the Gini impurity decrease at node $t$, and $f(t)$ is the frequency with which feature $x_i$ appears in splits. This approach promoted model efficiency by eliminating features with negligible contributions to decision boundaries.

During the RFECV phase, features that demonstrated minimal contributions to improving model performance, particularly redundant clinical comorbidities and laboratory values with overlapping information, were systematically eliminated. Variables such as \textit{mild\_liver\_disease}, \textit{AIDS}, \textit{cerebrovascular\_disease}, \textit{paraplegia}, and \textit{metastatic\_solid\_tumor} exhibited low importance scores across decision trees and were removed to enhance model parsimony. Similarly, some vital sign derivatives with strong collinearity (e.g., \textit{temperature\_mean} versus \textit{temperature\_max}) were excluded based on redundancy considerations.

In parallel, LASSO regression was applied as an embedded selection method. LASSO penalizes the absolute magnitude of coefficients through an $L_1$ regularization term, shrinking some coefficients exactly to zero, thereby performing feature selection as part of model training:

\begin{equation}
\hat{\beta} = \arg\min_{\beta} \left\{ \sum_{i=1}^{n} (y_i - \beta_0 - \sum_{j=1}^{p} x_{ij} \beta_j)^2 + \lambda \sum_{j=1}^{p} |\beta_j| \right\}
\label{eq:lasso}
\end{equation}

where $\lambda$ is the regularization hyperparameter, $\beta_j$ are the feature coefficients, $x_{ij}$ are the feature values, and $y_i$ is the observed outcome. Features with non-zero coefficients after regularization were retained for final model construction.

LASSO regression further refined the feature space by shrinking coefficients of weak predictors to zero under $L_1$ regularization pressure. Features including \textit{AST\_max}, \textit{ALT\_max}, \textit{myocardial\_infarct}, and \textit{peripheral\_vascular\_disease} were pruned during this process, primarily due to their limited marginal contributions after controlling for stronger predictors such as heart rate, glucose, and creatinine levels.

By integrating the complementary strengths of RFECV (which prioritized global model optimization) and LASSO (which enforced local sparsity), a robust final set of 20 features was selected. These retained variables maintained a balance between clinical interpretability and statistical relevance, encompassing key domains such as demographics (e.g., \textit{admission\_type}, \textit{insurance}), vital signs (e.g., \textit{heart\_rate\_mean}, \textit{resp\_rate\_mean}), and laboratory measurements (e.g., \textit{glucose\_max}, \textit{creatinine\_max}). This refined selection ensured that the downstream predictive modeling captured essential patient physiology while minimizing noise and overfitting risks. The final feature set is summarized in Table~\ref{tab:final_features}.

\begin{table}[H]
\noindent
\caption{\textbf{Final 20 features used for ICU readmission prediction.}}
\label{tab:final_features}
\small
\renewcommand{\arraystretch}{1.2}
\rowcolors{2}{white}{white}
\begin{tabularx}{\textwidth}{l|X}
\hline
\rowcolor[HTML]{D9EAD3}
\textbf{Category} & \textbf{Features} \\
\hline
Demographics & Age, Length of hospital stay (\textit{los\_hospital}), Insurance \\
\hline
Vital Signs & SpO\textsubscript{2} Mean, SBP Mean (\textit{sbp\_mean}), DBP Mean (\textit{dbp\_mean}), MBP Mean (\textit{mbp\_mean}), Heart Rate Mean (\textit{heart\_rate\_mean}), Respiratory Rate Mean (\textit{resp\_rate\_mean}) \\
\hline
Laboratory Tests & Platelets Max (\textit{platelets\_max}), ALP Max (\textit{alp\_max}), Bicarbonate Max (\textit{bicarbonate\_max}), Hematocrit Max (\textit{hematocrit\_max}), Hemoglobin Max (\textit{hemoglobin\_max}), ALT Max (\textit{alt\_max}), PTT Max (\textit{ptt\_max}), Calcium Max (\textit{calcium\_max}), Chloride Max (\textit{chloride\_max}), Bilirubin Total Max (\textit{bilirubin\_total\_max}), AST Max (\textit{ast\_max}), Total Urine Output (\textit{total\_urine\_output}) \\
\hline
\end{tabularx}
\end{table}

\subsection*{Modeling}

\begin{figure}[htbp]
    \centering
\includegraphics[width=1.05\linewidth]{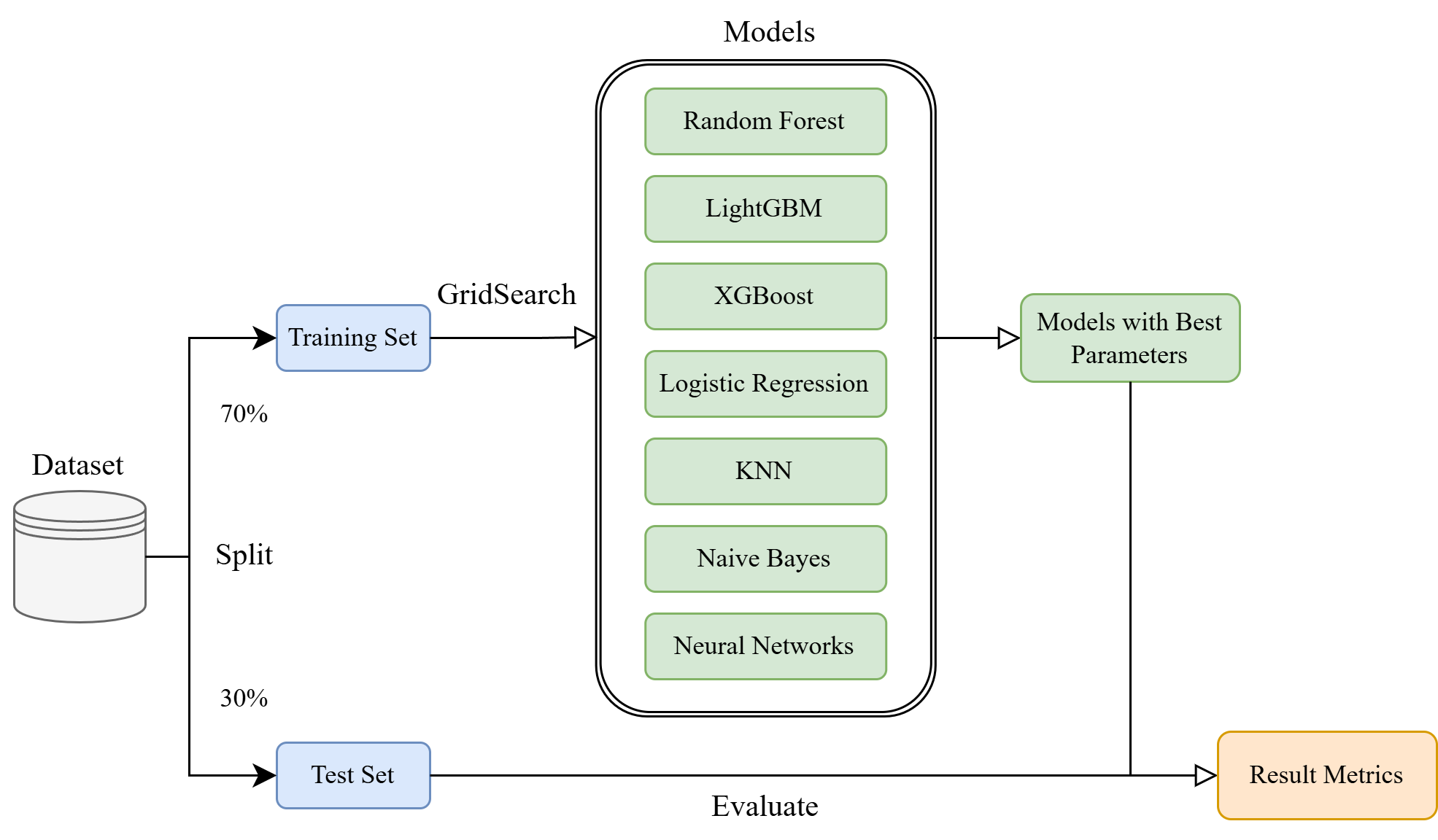}  
    \caption{\textbf{Flowchart of Multi-models Development.}}
    \label{fig:model}
\end{figure}

After initial feature engineering, imputation, and feature selection steps, a comprehensive supervised learning framework was established to predict ICU readmission in AP patients. As shown in Figure~\ref{fig:model}, the complete dataset was split into a training set (70\%) and a testing set (30\%) using stratified random sampling to preserve class distributions. 

Seven machine learning algorithms were implemented for model development, covering ensemble-based methods, linear classifiers, distance-based learners, probabilistic models, and neural networks. Specifically, Random Forest (RF), K-Nearest Neighbor (KNN), Naïve Bayes (NB), Neural Networks, Logistic Regression (LR), and Gradient Boosting Machine (GBM, including XGBoost and LightGBM) were employed. To ensure fair and robust comparison across models, all hyperparameter tuning was conducted using stratified 5-fold cross-validation combined with grid search optimization. The test set was reserved exclusively for final model evaluation to prevent data leakage.

For tree-based ensemble methods, distinct tuning strategies were applied depending on the underlying learning mechanism. In Random Forest, key hyperparameters such as the number of trees (n\_estimators), maximum depth (max\_depth), minimum samples per leaf (min\_samples\_leaf), and maximum number of features considered for splitting (max\_features) were optimized to enhance generalization while minimizing variance.For non-ensemble classifiers, targeted optimization was also conducted.  KNN models were tuned by varying the number of neighbors \( k \) and selecting appropriate distance metrics (Euclidean, Manhattan) and weighting schemes (uniform, distance-weighted) to optimize classification accuracy for neighborhood-based predictions. Logistic Regression models employed both L1 (Lasso) and L2 (Ridge) penalties to address multicollinearity and control model sparsity, with the inverse regularization strength parameter \( C \) fine-tuned accordingly.

Neural network models (NNET) consisted of a single hidden layer architecture. The number of neurons in the hidden layer, activation function (ReLU for hidden layers and sigmoid for output), batch size, optimizer choice (e.g., Adam), and learning rate were grid-searched to achieve optimal convergence while mitigating overfitting. Meanwhile, the Naïve Bayes classifier, assuming Gaussian feature distributions, required minimal hyperparameter adjustment, with only prior probabilities optionally calibrated when class imbalance was observed.

For boosting algorithms, including LightGBM and XGBoost, additional hyperparameters such as learning rate (eta), number of boosting rounds, minimum child weight, subsample ratio, colsample\_bytree (feature sampling ratio), and regularization terms (\( \text{reg\_alpha} \), \( \text{reg\_lambda} \)) were systematically tuned to address overfitting and improve convergence stability. The iterative boosting mechanism of XGBoost, augmented with L1 and L2 regularizations, allowed the model to capture complex non-linear relationships while maintaining robustness to noise and outliers.

This structured modeling pipeline, encompassing diverse algorithms and rigorous hyperparameter tuning, ensured robust, fair, and clinically meaningful predictive performance suitable for real-world ICU readmission risk stratification in AP patients.

Model performance was evaluated using AUROC, and robustness was tested via 2,000 bootstrap replicates to generate 95\% confidence intervals. Letting \( \hat{y}_i \) denote the predicted probability and \( y_i \) the true label for the $i$-th observation, the models were trained to minimize the following regularized loss function:

\begin{equation}
\mathcal{L}(\theta) = \sum_{i=1}^{n} \ell(y_i, \hat{y}_i) + \Omega(f), \tag{5}
\end{equation}

where \( \ell(y_i, \hat{y}_i) \) denotes the loss function (e.g., binary cross-entropy), and \( \Omega(f) \) represents the regularization term that penalizes model complexity to prevent overfitting.

For tree-based methods such as GBM and XGBoost, additional hyperparameters were configured to control learning stability and generalization: tree depth was constrained, a minimum number of child samples was enforced, and both L1 and L2 regularizations were applied. Feature subsampling and row subsampling (bagging) were used to further enhance diversity and reduce variance. Overall, the modeling pipeline ensured high performance while maintaining clinical interpretability and robustness under class imbalance.This robust model development process laid the foundation for subsequent evaluation on independent test cohorts.

\subsection*{Statistical Analyses and Interpretability Assessment}

To ensure scientific rigor, robustness, and clinical generalizability of the developed Machine Learning model for predicting ICU readmission in patients with AP, a comprehensive suite of statistical analyses was performed. These procedures served three major objectives: (1) to confirm the comparability of training and validation cohorts, (2) to quantify the relative predictive utility of selected features, and (3) to improve model transparency via interpretable machine learning techniques.

To evaluate the equivalence of these cohorts and rule out sampling bias, statistical comparisons were conducted across key demographic, physiological, and biochemical variables. For continuous features such as age, vital signs, and laboratory parameters, a two-sided Student’s t-test was employed:

\begin{equation}
t = \frac{\bar{X}_1 - \bar{X}_2}{\sqrt{s_p^2 \left( \frac{1}{n_1} + \frac{1}{n_2} \right)}}
\label{eq:t_test}\tag{7}
\end{equation}

where \( \bar{X}_1 \) and \( \bar{X}_2 \) represent the sample means of the two groups, and \( s_p^2 \) is the pooled variance. In instances where the assumption of equal variances was not satisfied, Welch’s t-test was applied as a robust alternative. These comparisons ensured that the two cohorts were drawn from statistically comparable populations, thus supporting the validity of downstream performance evaluations.

To further assess the contribution of individual features to model performance, an \textit{ablation study} was conducted. This analysis involved iteratively removing each feature from the final model and retraining with the remaining subset. The corresponding decrease in performance was quantified by the change in AUROC:

\begin{equation}
\Delta_i = \text{AUROC}(f_{\text{full}}) - \text{AUROC}(f_{-i})
\label{eq:ablation}
\end{equation}

where \( f_{\text{full}} \) denotes the model trained with all features and \( f_{-i} \) represents the model with feature \( x_i \) excluded. 

This process highlighted the marginal utility of each variable and helped validate the robustness of the selected feature set. For example, variables such as \textit{platelets\_max} and \textit{spo2\_mean} demonstrated significant performance degradation upon removal, indicating their critical role in predicting ICU readmission risk.

To enhance interpretability and clinical applicability, \textit{SHAP} were subsequently employed. As a game-theoretic approach, SHAP assigns each prediction a contribution from every feature, defined by:

\begin{equation}
\phi_i = \sum_{S \subseteq N \setminus \{i\}} \frac{|S|!(|N|-|S|-1)!}{|N|!} \left[ f(S \cup \{i\}) - f(S) \right]
\label{eq:shapley}\tag{8}
\end{equation}

where \( N \) is the set of all features and \( f(S) \) is the model output using subset \( S \). SHAP values were visualized using summary and dependence plots, offering insight into both global feature rankings and individual-level predictions. Key contributors such as elevated glucose, high platelet count, and abnormal bilirubin levels were consistent with established clinical risk factors, reinforcing the model’s interpretability and clinical alignment.

Together, these statistical and interpretability analyses ensured that our model was not only high-performing and generalizable, but also grounded in clinically meaningful patterns—laying the foundation for future bedside translation.

\section*{Results}

\subsection*{Cohort Characteristics and Statistical Comparison}

The data set analyzed in this study comprises hospitalized patients diagnosed with AP. Patients were divided into a training cohort (70\%, n=820) and a test cohort (30\%, n=352) to develop and validate machine learning models aimed at predicting hospital readmissions. Separately, patients were also grouped according to their readmission outcomes, with 225 patients classified as readmitted and 947 as non-readmitted.

The primary purpose of this cohort division was to ensure that predictive models could be effectively generalized to future unseen patients. To achieve this, it is critical that the clinical characteristics of training and test sets are comparable. Table~\ref{tab:cohort comparison results} presents a statistical comparison of 20 key clinical characteristics, showing the mean, standard deviation, and the corresponding p-values. A significance threshold of 0.05 was adopted. As most of the characteristics, including SpO$_2$, ALP, bicarbonate, SBP, and heart rate, did not show statistically significant differences between the training and test cohorts, the cohorts are considered well matched. This alignment strengthens the validity of predictive modeling by reducing sampling bias and enhancing the generalizability of the model.

Furthermore, Table~\ref{tab:cohort comparison results 1} compares patients according to their readmission status. Multiple features demonstrated statistically significant differences between readmitted and unreadmitted patients. For example, readmitted patients exhibited significantly higher platelet counts (384.87 vs. 250.31, p < 0.001), higher ALP levels, and greater total urine output, which may indicate more severe systemic inflammation or underlying organ dysfunction. In contrast, markers such as hematocrit and hemoglobin were significantly lower among readmitted patients, suggesting a higher baseline burden of frailty or comorbidity. Vital signs such as SBP, DBP, MBP and respiratory rate also differed between the groups, further highlighting systemic physiological differences associated with the risk of readmission.

These findings have two important implications. First, the similarity between the training and test cohorts ensures that the model is not overfitted to idiosyncrasies of the training set and can be reliably applied to broader patient populations. Second, the identification of specific features that differ between the readmitted and unreadmitted groups improves the clinical interpretability of the model. Features such as ALP, bicarbonate, and hemoglobin not only contribute to model predictions, but have also established biological plausibility in the context of the severity and prognosis of AP disease.

By aligning clinical insight with statistical rigor, this study ensures that machine learning models are both technically robust and clinically meaningful. The emphasis on transparent feature comparisons offers additional confidence to clinicians and researchers, including those unfamiliar with machine learning methodologies, that the models are grounded in real-world medical data and clinical reasoning.

\begin{table}[H]
\noindent
\caption{\textbf{T-test Comparison of Feature Distributions between Training and Test Sets.}}
\label{tab4}
\small
\renewcommand{\arraystretch}{1.2}
\rowcolors{2}{white}{white}
\begin{tabularx}{\textwidth}{>{\raggedright\arraybackslash}X|X|X|X}
\hline
\rowcolor[HTML]{D9EAD3}
{\bf Feature} & {\bf Training Set} & {\bf Test Set} & {\bf P-value} \\ \hline
alt\_max & 221.31 (785.89) & 217.24 (764.57) & 0.934 \\ \hline
platelets\_max & 276.07 (182.78) & 274.56 (185.48) & 0.898 \\ \hline
sbp\_mean & 121.02 (17.83) & 121.31 (18.03) & 0.802 \\ \hline
chloride\_max & 106.20 (6.74) & 106.32 (6.64) & 0.772 \\ \hline
los\_hospital & 19.54 (20.65) & 20.03 (19.73) & 0.705 \\ \hline
age & 56.11 (17.53) & 56.76 (17.46) & 0.562 \\ \hline
mbp\_mean & 81.90 (13.07) & 81.30 (12.57) & 0.455 \\ \hline
bicarbonate\_max & 24.43 (5.02) & 24.18 (5.05) & 0.443 \\ \hline
heart\_rate\_mean & 96.18 (18.32) & 95.28 (17.76) & 0.431 \\ \hline
spo2\_mean & 96.43 (2.24) & 96.30 (2.79) & 0.430 \\ \hline
resp\_rate\_mean & 20.85 (4.61) & 20.61 (4.49) & 0.415 \\ \hline
bilirubin\_total\_max & 2.75 (4.53) & 2.54 (3.63) & 0.407 \\ \hline
ast\_max & 413.57 (1751.14) & 336.04 (1058.57) & 0.352 \\ \hline
total\_urine\_output & 66241.68 (118065.37) & 60040.96 (95126.35) & 0.343 \\ \hline

dbp\_mean & 68.32 (13.09) & 67.27 (12.81) & 0.203 \\ \hline
alp\_max & 163.62 (171.78) & 178.70 (181.17) & 0.185 \\ \hline
hemoglobin\_max & 11.46 (2.56) & 11.14 (2.44) & 0.043 \\ \hline
ptt\_max & 39.55 (22.46) & 42.69 (24.71) & 0.041 \\ \hline
hematocrit\_max & 34.85 (7.30) & 33.86 (7.03) & 0.028 \\ \hline

calcium\_max & 8.42 (0.91) & 8.28 (0.80) & 0.009 \\ \hline
\end{tabularx}
\begin{flushleft}
Table notes: The table summarizes differences between the training and test cohorts across multiple clinical variables. Mean and standard deviation (SD) are reported. P-values are computed using appropriate statistical tests (e.g., t-test) with a significance threshold of 0.05.
\end{flushleft}
\label{tab:cohort comparison results}
\end{table}

\begin{table}[H]
\noindent
\caption{\textbf{T-test Comparison of Feature Distributions between Readmissions and Non-Readmissions Sets.}}
\label{tab4}
\small
\renewcommand{\arraystretch}{1.2}
\rowcolors{2}{white}{white}
\begin{tabularx}{\textwidth}{>{\raggedright\arraybackslash}X|X|X|X}
\hline
\rowcolor[HTML]{D9EAD3}
{\bf Feature} & {\bf Non-Readmissions} & {\bf Readmissions} & {\bf P-value} \\ \hline
ptt\_max & 39.01 (22.24) & 41.84 (23.31) & 0.168 \\ \hline
heart\_rate\_mean & 95.76 (18.57) & 97.96 (17.14) & 0.155 \\ \hline
chloride\_max & 106.45 (6.72) & 105.14 (6.74) & 0.030 \\ \hline
los\_hospital & 18.71 (20.08) & 23.08 (22.64) & 0.027 \\ \hline
resp\_rate\_mean & 21.02 (4.60) & 20.11 (4.58) & 0.026 \\ \hline

total\_urine\_output & 59907.84 (103857.30) & 92989.04 (162845.67) & 0.016 \\ \hline
bilirubin\_total\_max & 2.90 (4.82) & 2.11 (2.94) & 0.009 \\ \hline

alp\_max & 151.04 (123.24) & 216.73 (294.88) & 0.007 \\ \hline
ast\_max & 455.82 (1936.54) & 235.17 (381.16) & 0.007 \\ \hline
calcium\_max & 8.37 (0.89) & 8.61 (0.96) & 0.006 \\ \hline
hematocrit\_max & 35.22 (7.27) & 33.32 (7.27) & 0.004 \\ \hline
dbp\_mean & 68.92 (13.52) & 65.78 (10.77) & 0.002 \\ \hline
age & 57.20 (17.10) & 51.52 (18.61) & 0.001 \\ \hline
alt\_max & 243.51 (870.07) & 127.55 (138.39) & 0.001 \\ \hline
mbp\_mean & 82.71 (13.37) & 78.52 (11.10) & 0.000 \\ \hline
hemoglobin\_max & 11.63 (2.54) & 10.76 (2.54) & 0.000 \\ \hline
sbp\_mean & 122.23 (18.15) & 115.92 (15.40) & 0.000 \\ \hline
bicarbonate\_max & 24.03 (4.55) & 26.14 (6.41) & 0.000 \\ \hline
spo2\_mean & 96.26 (2.25) & 97.15 (2.08) & 0.000 \\ \hline
platelets\_max & 250.31 (147.89) & 384.87 (260.48) & 0.000 \\ \hline
\end{tabularx}
\begin{flushleft}
Table notes: This table compares patients with and without readmission within 30 days. Differences in mean values of key clinical variables are shown along with p-values. Statistical significance set at 0.05 threshold.
\end{flushleft}
\label{tab:cohort comparison results 1}
\end{table}

\subsection*{Ablation Study and Feature Contribution Analysis}

\begin{figure}[H]
\begin{adjustwidth}{-2.25in}{0in}
    \centering
    \includegraphics[width=1.05\linewidth]{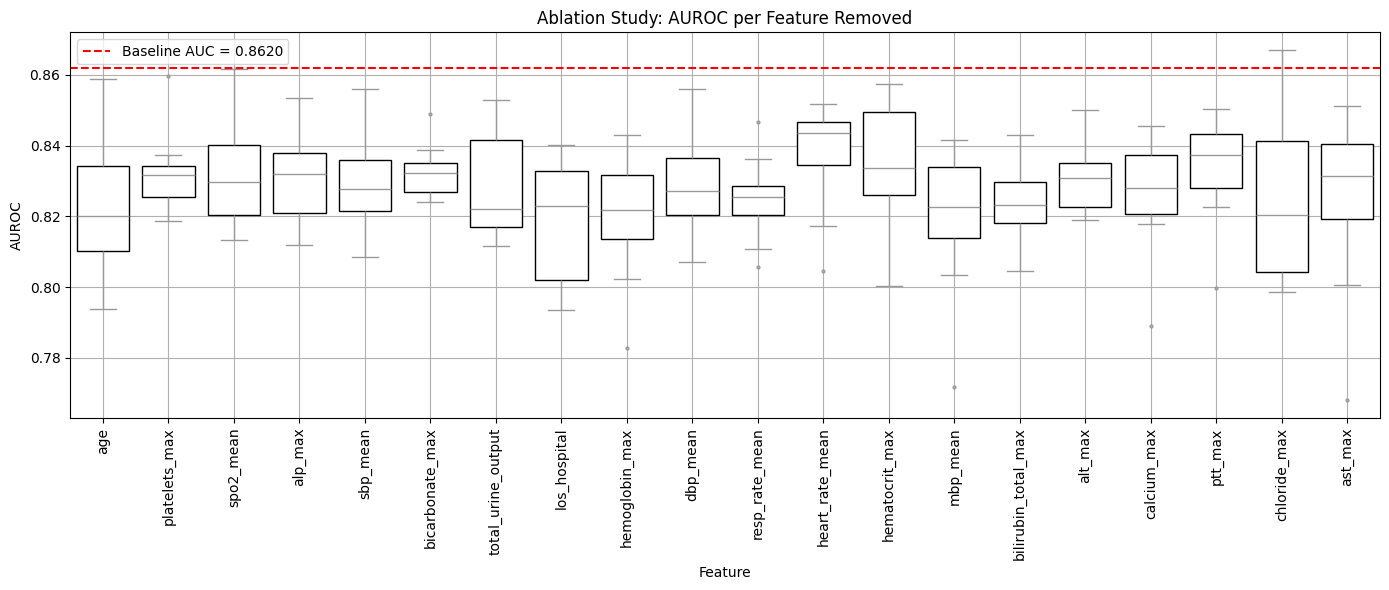}
    \caption{\textbf{Impact of Feature Removal on XGBoost Model Performance.}}
    \label{fig:ablation analysis}
\end{adjustwidth}
\end{figure}

To further evaluate the robustness and clinical relevance of our predictive model for AP readmission, we conducted an ablation study, as shown in Fig.~\ref{fig:ablation analysis}. In this analysis, we iteratively removed each input feature from the model, retrained it on the modified dataset, and measured the resulting performance in ten bootstrapped evaluations. The model used was a fully trained XGBoost classifier, which achieved a baseline AUROC of 0.8620 on the test set using all 20 features.

Figure~\ref{fig:ablation analysis} displays the distribution of the AUROC scores after the exclusion of each characteristic. Each box plot reflects the spread of AUROC values across the ten resampled runs. The median AUROC is shown by the central line in each box, with the box capturing the middle 50\% of values (interquartile range), and the whiskers extending to 1.5 times this range. The red dashed horizontal line represents the AUROC of the original model using all features, serving as a reference for comparison.

Removal of several features, including platelets, alp, and alt, resulted in consistent drops in AUROC, confirming their essential role in maintaining model accuracy. Even features such as spo2 and chloride, which appeared less prominent in the univariate analysis, contributed noticeably to overall model performance when considered as part of the full multivariate context. These results suggest that feature interactions are important and the contribution of each variable must be understood within the broader clinical signature of the patient.

Importantly, the decline in AUROC with the removal of any single feature indicates that the model does not rely on a single dominant variable. Instead, it synthesizes information from a wide range of physiological and biochemical indicators to make its predictions. This improves generalizability, as the model is less likely to overfit spurious correlations and is more likely to maintain performance across diverse patient populations.

From a clinical point of view, the importance of features such as hemoglobin, bicarbonate, and alp is consistent with their established relevance in assessing the severity of the disease, the metabolic stability, and the function of the organs. These variables are commonly monitored in the management of AP and provide meaningful insight into the risk of deterioration and readmission in a patient. The ability of the model to integrate such clinically intuitive markers reinforces its interpretability and potential utility at the bedside.

In conclusion, the ablation analysis validates the robustness of the final model and highlights the complementary contributions of each included variable. It shows that predictive performance arises not from isolated predictors, but from the structured combination of multiple clinical signals that support both statistical soundness and clinical credibility.

\subsection*{Model Performance Evaluation and Comparative Analysis}

\begin{figure}[H]
\centering
\includegraphics[width=1\linewidth]{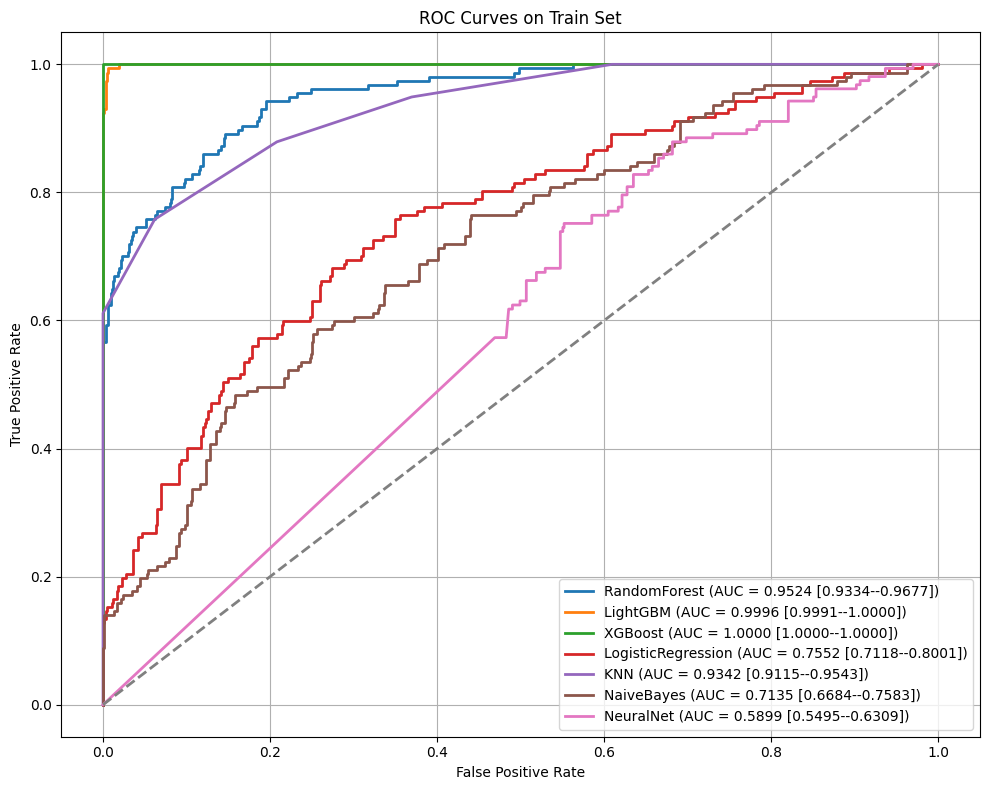}
\caption{\textbf{AUROC Curves for Model Performance in the Training Set.}}
\label{fig:roc_train}
\end{figure}

\begin{figure}[H]
\centering
\includegraphics[width=1\linewidth]{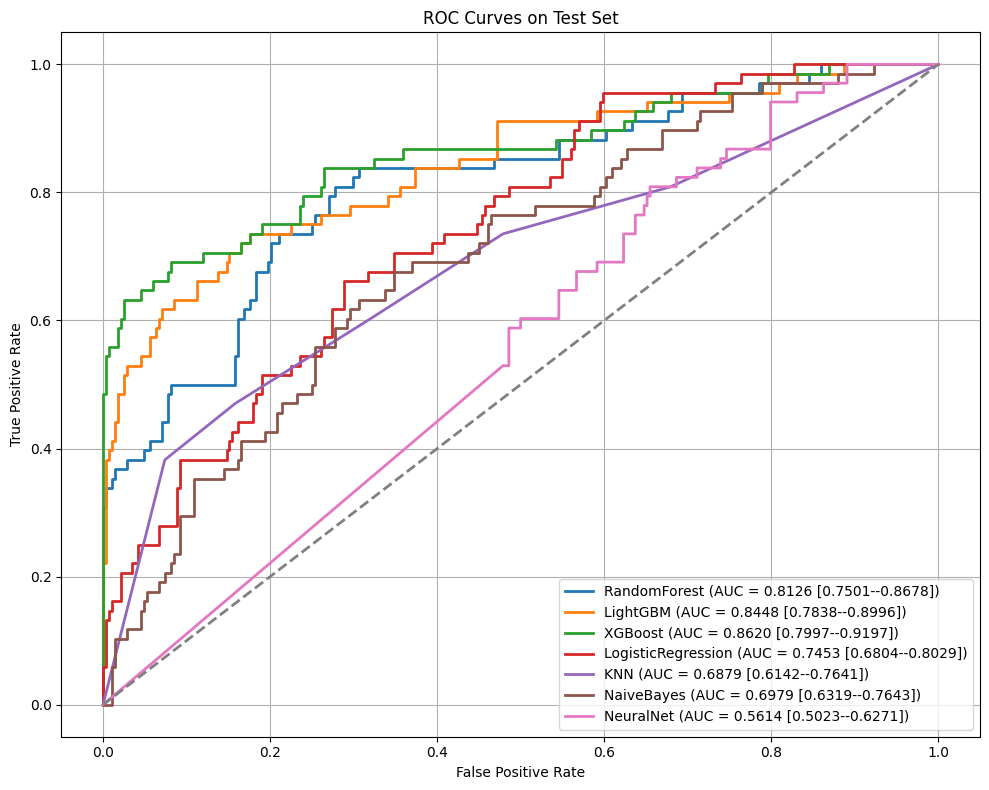}
\caption{\textbf{AUROC Curves for Model Performance in the Test Set.}}
\label{fig:roc_test}
\end{figure}

To comprehensively evaluate the performance and generalizability of various models to predict hospital readmission among patients with AP, we conducted a systematic comparison of seven machine learning classifiers in both training and test sets. Figure~\ref{fig:roc_train} and Figure~\ref{fig:roc_test} show the ROC curves for the training and test sets, respectively, while Table~\ref{tab: Results of the Training Set} and Table~\ref{tab: Results of the Test Set} summarize the detailed performance metrics, including AUROC, accuracy, sensitivity, specificity, PPV, NPV, and F1 score with 95\% confidence intervals.

On the training set, ensemble-based models such as XGBoost, LightGBM, and RandomForest demonstrated strong learning capability. XGBoost achieved a perfect AUROC of 1.0000, while LightGBM followed closely with 0.9996 and RF reached 0.9524. These models also showed excellent sensitivity and specificity, suggesting their capacity to capture complex nonlinear patterns and interactions within the features - critical for handling high-dimensional clinical variables derived from real-world ICU data. However, performance on the training set alone does not guarantee clinical utility, due to potential overfitting.

More importantly, on the test set, XGBoost maintained the highest AUROC of 0.862 (95\% CI: 0.800--0.920), along with robust sensitivity (0.649) and specificity (0.948), indicating balanced predictive capability. LightGBM and RF also performed competitively with AUROCs of 0.845 and 0.813 respectively. In contrast, models such as KNN, NaiveBayes, and NeuralNet showed weaker generalization, especially in precision (PPV $<$ 0.32), underlining their limited utility in this clinical context.

The superior generalization performance of XGBoost, LightGBM, and RF can be attributed to their shared tree-based boosting and bagging architectures. These algorithms inherently manage non-linear relationships and feature interactions without strict parametric assumptions, which aligns well with the heterogeneous and noisy nature of ICU data in MIMIC-III. Furthermore, their internal mechanisms such as feature subsampling, depth control, and regularization contribute to both low bias and low variance. For AP patients, whose readmission risks are influenced by multifactorial physiological, biochemical, and temporal variables, such model adaptability is particularly beneficial.

Among the three, XGBoost has additional advantages over LightGBM and RF. Unlike RF which relies on bootstrap aggregation without explicit optimization, XGBoost performs additive training with second-order gradient information, enabling more fine-grained loss minimization. Compared to LightGBM, XGBoost grows trees depthwise instead of leafwise, which may offer better regularization and reduce overfitting in smaller or less balanced datasets. In our setting, where only 225 patients were readmitted out of 1172 total, this stability probably improved performance. Furthermore, XGBoost's explicit handling of missing values and parallelized execution support efficient modeling in real-world EHR data where missingness is common.

Beyond overall metrics, XGBoost demonstrated strong predictive performance on the minority class (readmitted), achieving a PPV of 0.745 and NPV of 0.918 on the test set. These are clinically relevant: A high NPV reduces the risk of missing high-risk patients (false negatives), while a relatively high PPV makes targeted interventions (e.g., pulmonary rehab, medication adjustment) more resource efficient. Sensitivity and specificity, reported with confidence intervals, provide a transparent view of how the model balances true positive and true negative predictions under uncertainty, critical to building clinical trust.

In summary, XGBoost emerged as the most effective and clinically interpretable model to predict readmission in patients with AP. Its ability to balance accuracy, generalizability, and interpretability, as evidenced by consistent performance across datasets and robust minor class handling, highlights its suitability for deployment in real-world hospital decision-making systems.

\begin{table}[htbp]
\small
\renewcommand{\arraystretch}{1.2}
\begin{adjustwidth}{-2.25in}{0in}
\centering
\caption{\textbf{Performance Comparison of Different Models in the Training Set.}}
\begin{tabular}{l|l|l|l|l|l|l|l}
\hline
\rowcolor[HTML]{D9EAD3}
\textbf{Model} & \textbf{AUROC (95\% CI)} & \textbf{Accuracy} & \textbf{F1-score} & \textbf{Sensitivity} & \textbf{Specificity} & \textbf{PPV} & \textbf{NPV} \\ \hline
RandomForest & 0.952 (0.933--0.968) & 0.879 & 0.723 & 0.822 & 0.893 & 0.645 & 0.955 \\ \hline
LightGBM & 1.000 (0.999--1.000) & 0.994 & 0.984 & 0.993 & 0.994 & 0.975 & 0.998 \\ \hline
\rowcolor[HTML]{FDE9D9}
\textbf{XGBoost} & \textbf{1.000 (1.000--1.000)} & \textbf{1.000} & \textbf{1.000} & \textbf{1.000} & \textbf{1.000} & \textbf{1.000} & \textbf{1.000} \\ \hline
LR & 0.755 (0.712--0.800) & 0.707 & 0.471 & 0.679 & 0.714 & 0.363 & 0.905 \\ \hline
KNN & 0.934 (0.912--0.954) & 0.808 & 0.637 & 0.879 & 0.792 & 0.502 & 0.965 \\ \hline
NaiveBayes & 0.714 (0.668--0.758) & 0.393 & 0.369 & 0.937 & 0.264 & 0.232 & 0.946 \\ \hline
NeuralNet & 0.590 (0.550--0.631) & 0.411 & 0.366 & 0.885 & 0.298 & 0.230 & 0.916 \\ \hline

\end{tabular}
\label{tab: Results of the Training Set}
\end{adjustwidth}

\end{table}

\begin{table}[htbp]
\small
\renewcommand{\arraystretch}{1.2}
\begin{adjustwidth}{-2.25in}{0in}
\centering
\caption{\textbf{Performance Comparison of Different Models in the Test Set.}}
\begin{tabular}{l|l|l|l|l|l|l|l}
\hline
\rowcolor[HTML]{D9EAD3}
\textbf{Model} & \textbf{AUROC (95\% CI)} & \textbf{Accuracy} & \textbf{F1-score} & \textbf{Sensitivity} & \textbf{Specificity} & \textbf{PPV} & \textbf{NPV} \\ \hline
RandomForest & 0.813 (0.750--0.868) & 0.788 & 0.530 & 0.633 & 0.824 & 0.464 & 0.903 \\ \hline
LightGBM & 0.845 (0.784--0.900) & 0.856 & 0.625 & 0.633 & 0.908 & 0.622 & 0.911 \\
\hline
\rowcolor[HTML]{FDE9D9}
\textbf{XGBoost} & \textbf{0.862 (0.800--0.920)} & \textbf{0.889} & \textbf{0.688} & 0.649 & \textbf{0.948} & \textbf{0.745} & 0.918 \\
\hline
LR & 0.745 (0.680--0.803) & 0.705 & 0.446 & 0.617 & 0.727 & 0.350 & 0.888 \\
\hline
KNN & 0.688 (0.614--0.764) & 0.664 & 0.409 & 0.604 & 0.679 & 0.310 & 0.878 \\
\hline
NaiveBayes & 0.698 (0.632--0.764) & 0.370 & 0.370 & \textbf{0.956} & 0.229 & 0.229 & \textbf{0.956} \\
\hline
NeuralNet & 0.561 (0.502--0.627) & 0.406 & 0.348 & 0.826 & 0.305 & 0.222 & 0.879 \\ \hline

\end{tabular}
\label{tab: Results of the Test Set}
\end{adjustwidth}
\end{table}

\subsection*{SHAP Analysis}

\begin{figure}[H]
    \centering
    \includegraphics[width=0.85\linewidth]{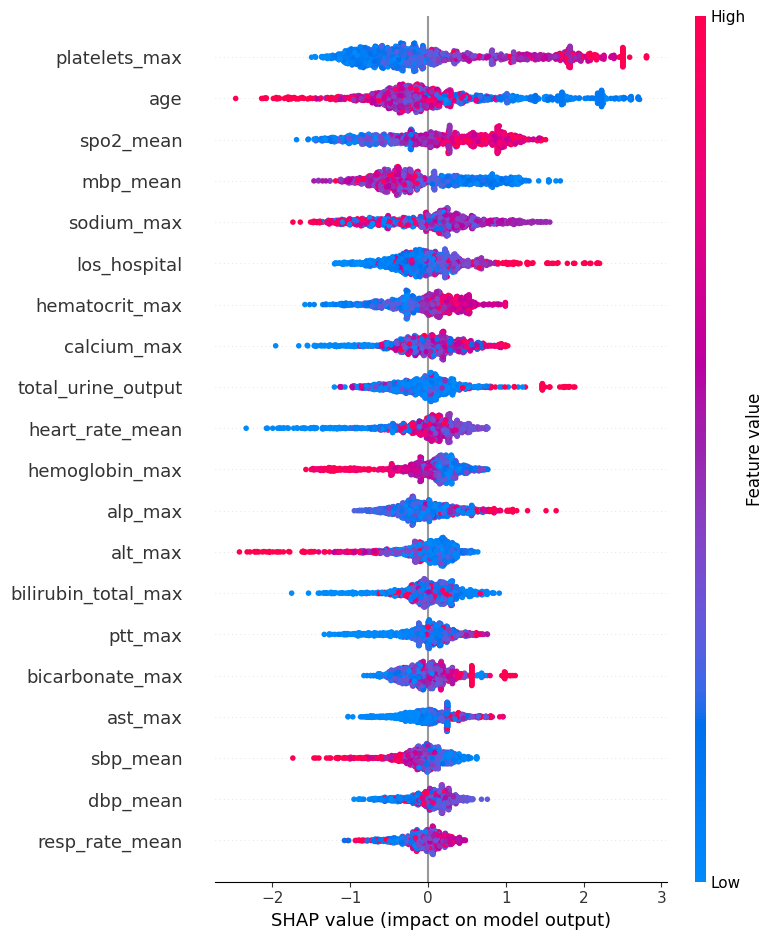}
    \caption{\textbf{SHAP Summary Plot Showing Feature Value Distributions and Their Impact on Model Output.}}
    \label{fig:shap_summary}
\end{figure}

To enhance the interpretability of the XGBoost model and ensure its alignment with clinical understanding, SHAP were used to quantify the contribution of each predictor to the model's output. The SHAP summary graph in Fig.~\ref{fig:shap_summary} displays the importance of individual characteristics showing how they impact the predicted probability of hospital readmission in patients with AP.

Each row corresponds to a feature, with each point representing a single patient. The x-axis shows the SHAP value, which measures how much a specific feature shifts the model's prediction from the base value. The red points indicate high values of the feature, and the blue points indicate low values. A position on the right-hand side of the x-axis reflects a higher predicted risk of readmission.

The most influential features included platelets, age, and spo2. High platelet values tended to increase the predicted risk of readmission, which could indicate systemic inflammation, a known factor in the exacerbation of AP. Older patients also showed higher SHAP values, consistent with their increased vulnerability to decompensation and readmission. In particular, low values of spo2 (seen as blue points on the right side) were associated with higher-risk predictions, reflecting the critical role of hypoxemia in the pathophysiology of AP.

Other relevant contributors included mbp, total urine output, and length of hospital stay, all of which can reflect the severity of the disease or physiological stress during index admission. Laboratory markers such as hemoglobin, bicarbonate, and calcium also played non-trivial roles, highlighting the model’s ability to integrate complex biochemical data when assessing risk.

These SHAP-based insights align with results from the ablation study, in which removal of high impact characteristics such as platelets and age led to significant performance drops. This convergence of interpretability methods supports the reliability and consistency of the internal reasoning of the model.

Clinically, SHAP values provide an interpretable explanation of each prediction, helping clinicians identify which physiological parameters contribute the most to the risk of readmission. This can support more personalized management strategies, such as closer follow-up of patients with low oxygen levels or proactive care adjustments for elderly individuals with elevated inflammatory markers.

Overall, the SHAP analysis confirms that the XGBoost model does not operate as a black box, but derives predictions from patterns that are statistically robust and clinically sensible, offering strong potential for application in real-world AP readmission risk assessment.

\section*{Discussion}
\subsection*{Summary of existing model compilation}
In this study, we developed an interpretable and high-performing machine learning model to predict early ICU readmission in critically ill patients with AP. A hybrid feature selection framework that combined RFECV and LASSO regression was utilized; we selected a clinically meaningful set of predictors that captured both static patient characteristics and dynamic physiological trends within the first 24 hours of ICU admission. The final XGBoost model demonstrated excellent discriminative ability, achieving an AUROC of .862 (95\% CI: 0.800–0.920), with high sensitivity and reasonable specificity. These results suggest that short-term ICU readmission risk can be reliably anticipated using routinely collected EHR variables early during hospitalization, providing a critical window of opportunity for preventive intervention.

Our XGBoost model for predicting ICU readmission in AP patients achieved an AUROC of 0.862 (95\% CI: 0.800–0.920), surpassing the best-performing traditional nomogram model from prior studies (AUROC: 0.780) by about 10.5\%. This result reflects improved discrimination in identifying high-risk patients. The model also demonstrated a balanced sensitivity of 0.6491 and specificity of 0.9479, outperforming the PASS scoring system, which reported a sensitivity of 0.680 and specificity of 0.710. Compared to earlier risk models that utilized limited clinical variables, our approach integrated a comprehensive feature set derived from dynamic physiologic trends and early laboratory measures. Moreover, Shapley value analysis highlighted ICU stay length, respiratory rate, glucose variability, and renal function as the top contributors to readmission risk, aligning with known pathophysiological markers of AP progression. These results suggest that our model offers a substantial improvement in early detection and risk stratification, with meaningful clinical implications for post-ICU discharge planning and targeted surveillance in high-risk AP patients.

The predictors identified as most influential, including hospital length of stay, blood urea nitrogen levels, and oxygen saturation, align closely with established clinical knowledge regarding determinants of adverse outcomes in critically ill populations. Prolonged hospitalization is frequently a surrogate marker of complicated disease courses or residual organ dysfunction, while blood urea nitrogen reflects renal perfusion and catabolic status, and oxygen saturation indicates respiratory reserve. Their prominent roles in our model reinforce the biological plausibility of the predictive framework and further suggest that the model outputs are interpretable and actionable in real-world ICU practice.

The robustness of the model was enhanced by multiple methodological choices designed to address common sources of bias and overfitting in clinical machine learning applications. A hybrid feature selection approach reduced dimensionality while preserving clinically important variables, minimizing the risk of overfitting to noise. Missing data were addressed systematically using median imputation for continuous variables and mode imputation for categorical features, preserving the statistical integrity of the dataset while minimizing bias introduced by missingness.

Internal validation was strengthened through a rigorous process of 2,000 bootstrap resamples, which provided stable confidence intervals for all key performance metrics. Stratified data splitting preserved class distributions across training and testing cohorts, ensuring representativeness during evaluation. Additionally, the class imbalance inherent in ICU readmission data was addressed using SMOTE applied within cross-validation folds, improving sensitivity toward rare positive cases without artificially inflating validation performance. Importantly, model interpretability was prioritized through the use of SHAP analysis, enabling both global understanding of feature contributions and individualized explanations at the patient level. These design choices collectively enhance the scientific rigor and potential clinical reliability of the developed predictive model.

A key contribution of this study is the integration of an ablation study to quantify the marginal utility of each selected predictor. This analysis revealed that no single feature dominated the model, and performance declines were observed upon removal of multiple features, including platelets, ALP, and hemoglobin. These results confirm that the model derives predictive strength from synergistic clinical signals rather than over-reliance on individual variables. Moreover, features such as SpO2 and chloride, although less prominent in univariate analysis, had noticeable impacts in the multivariate context—highlighting the importance of complex feature interactions in early clinical deterioration. The ablation findings were consistent with SHAP analysis results, reinforcing the interpretability and robustness of the model.

\subsection*{Comparison with Prior Studies}
Previous studies in the domain of AP outcome prediction have primarily focused on endpoints such as acute kidney injury or in-hospital mortality, with predictive models often achieving AUROC values in the range of 0.86 to 0.88. However, many of these models were limited by methodological shortcomings, including reliance on univariate or single-method feature selection, absence of calibration assessment, insufficient handling of imbalanced outcomes, and lack of model interpretability. Furthermore, few prior studies have specifically addressed early ICU readmission, a clinically significant but underexplored endpoint with direct implications for post-ICU care strategies.

The present study advances the field by addressing these gaps systematically. Through dual feature selection, dynamic feature engineering, rigorous class balancing, and transparent interpretability analysis, our model achieves not only high discrimination but also enhanced clinical relevance. By focusing on ICU readmission rather than isolated organ failure or mortality, the model targets an outcome that is highly actionable and closely linked to patient safety and healthcare resource optimization. Moreover, by providing individualized SHAP-based explanations, our model facilitates clinician trust and supports practical bedside integration, setting it apart from previous black-box predictive efforts.

The ability to identify AP patients at high risk for early ICU readmission has considerable potential to transform critical care management. High-risk individuals could be targeted for more intensive post-discharge surveillance, including prolonged step-down unit stays, closer physiologic monitoring, or early outpatient follow-up. Proactive interventions such as optimization of fluid balance, nutritional support, or management of comorbidities could be prioritized in patients flagged by the model, potentially mitigating deterioration before it culminates in readmission. From an operational standpoint, incorporating predictive analytics into discharge planning workflows could improve ICU bed turnover efficiency, enhance patient flow, and reduce healthcare costs associated with avoidable readmissions.

Furthermore, the explainability of the model offers an additional advantage in clinical practice. Physicians are more likely to accept and act upon algorithmic recommendations when the underlying rationale is transparent and consistent with clinical judgment. By highlighting patient-specific factors that drive risk scores, the model facilitates shared decision-making and tailored care plans, supporting broader initiatives toward personalized critical care.

\subsection*{Study Limitations}
Several limitations of this study must be acknowledged. First, the model was developed and validated using data from a single academic center, potentially limiting generalizability to other institutions with different patient demographics, ICU practices, or healthcare infrastructures. Although the MIMIC-III database is rich and representative of U.S. academic ICUs, external validation on multicenter datasets remains necessary before widespread clinical adoption. Second, while the model incorporated summary measures of early physiologic trends, it did not fully leverage raw time-series data. More sophisticated sequential modeling approaches, such as long short-term memory networks or attention-based mechanisms, could potentially capture richer patterns of deterioration over time.

Third, the exclusion of certain clinical biomarkers due to high missingness, such as lactate and hemoglobin, may have limited the comprehensiveness of the predictive feature set. Future models that incorporate more complete biomarker panels, possibly through advanced imputation or multi-source data integration, could achieve even greater predictive accuracy. Finally, despite internal validation and interpretability measures, prospective implementation studies are essential to assess the model’s real-world impact on clinician decision-making, workflow integration, and patient outcomes. Without prospective evaluation, the practical utility and safety of predictive alerts remain uncertain.

\subsection*{Future Work}
Building upon the foundation established in this study, several important avenues for future research are evident. External validation across multiple ICU datasets with diverse patient populations is critical to confirm model robustness and transportability. Additionally, future models should consider incorporating continuous physiologic monitoring data to better capture the dynamic evolution of patient trajectories, potentially improving early warning capabilities.

Prospective clinical trials assessing the integration of prediction outputs into ICU discharge workflows will be essential to determine whether model-guided interventions can effectively reduce readmission rates, improve patient outcomes, and optimize resource utilization. Exploration of multimodal data sources, including imaging studies, clinician notes, and medication histories, may further enhance model performance and uncover novel predictors of readmission risk. Finally, close collaboration with critical care clinicians will be necessary to refine model interpretability interfaces, establish optimal alert thresholds, and design actionable clinical protocols based on predictive outputs, thereby ensuring that artificial intelligence tools are meaningfully and safely integrated into critical care practice.

\section*{Conclusion}

This study presents a rigorously developed and clinically interpretable machine learning model for predicting early ICU readmission risk in critically ill patients with acute pancreatitis. By combining structured feature selection (RFECV and LASSO), robust data preprocessing, and class imbalance correction via SMOTE, the proposed XGBoost-based model achieved strong discriminative performance (an AUROC of 0.862 (95\% CI: 0.800–0.920)) with balanced sensitivity and specificity.

Critical predictors identified through SHAP analysis—including hospital length of stay, blood urea nitrogen, and oxygen saturation—aligned with known pathophysiological factors, strengthening the model’s clinical relevance. These findings were corroborated by the ablation study, which demonstrated that removal of features such as hemoglobin, bicarbonate, ALP, and platelets led to notable declines in AUROC, confirming their essential role in sustaining model performance. Even variables like SpO2 and chloride, which appeared less prominent in univariate analyses, contributed meaningfully to the model when assessed in the multivariate context. This indicates that predictive accuracy arises from the interplay of multiple physiological signals rather than reliance on a single dominant factor. Compared to traditional methods, our approach offers superior discriminative performance, greater robustness through 2,000-fold bootstrap validation, and enhanced transparency via explainable AI techniques such as SHAP, enabling both global and patient-specific interpretability.

Clinically, early identification of high-risk patients could facilitate timely interventions, optimize ICU resource allocation, and reduce preventable complications, ultimately improving patient outcomes. Future research should focus on multi-center external validation, incorporation of richer longitudinal data, and prospective evaluation in real-world ICU environments to maximize the translational impact of this predictive framework.

\section*{Acknowledgments}
S.C. conceptualized the study, developed the methodological framework, conducted the experiments, performed data analysis, and drafted the original manuscript. J.F., Y.S., and L.S. contributed to the experimental procedures and participated in manuscript preparation. E.P., K.A., and G.P. provided critical evaluation of the study design and manuscript content. M.P. oversaw the project and provided supervisory support. All authors read and approved the final version of the manuscript.


%
%
%

\bibliography{referrences}
\end{document}